\documentclass{article}
\usepackage{spconf,amsmath,amssymb,graphicx}
\usepackage{fancyhdr}

\usepackage{lipsum}

\title{COGNITIVE CODING OF SPEECH}

\name{Reza Lotfidereshgi, Philippe Gournay}
\address{Speech and Audio Research Group\\
Université de Sherbrooke\\
Sherbrooke (Québec) J1K 2R1 Canada}
\begin{document}
%
\maketitle
\begin{abstract}

We propose an approach for cognitive coding of speech by unsupervised extraction of contextual representations in two hierarchical levels of abstraction. Speech attributes such as phoneme identity that last one hundred milliseconds or less are captured in the lower level of abstraction, while speech attributes such as speaker identity and emotion that persist up to one second are captured in the higher level of abstraction. This decomposition is achieved by a two-stage neural network, with a lower and an upper stage operating at different time scales. Both stages are trained to predict the content of the signal in their respective latent spaces. A top-down pathway between stages further improves the predictive capability of the network. With an application in speech compression in mind, we investigate the effect of dimensionality reduction and low bitrate quantization on the extracted representations. The performance measured on the LibriSpeech and EmoV-DB datasets reaches, and for some speech attributes even exceeds, that of state-of-the-art approaches.

\end{abstract}
\section{Introduction}
\label{sec:intro}

The human cognitive system is known to have a hierarchical organization, the most cognitively complex operations being performed at the top of the hierarchy. While information mostly flows from the bottom to the top of the hierarchy, this bottom-up flow is often influenced by what is already known at the top of the hierarchy. Furthermore, there is substantial evidence for the predictive nature of this top-down influence \cite{bruner1957perceptual, heilbron2018great}. A parallel can be drawn between these defining elements of the cognitive system and the models used in machine learning. One of the first successful applications of deep learning was precisely in the field of automatic learning of hierarchical representations \cite{hinton2006reducing, vincent2008extracting}. It was also found that introducing top-down processes in hierarchical models improves the quality of learned representations, thereby increasing the accuracy of recognition systems based on these representations \cite{wang2019iterative, wen2018deep}. Predictive coding has also been shown to be a successful strategy in machine learning when processing various data modalities \cite{wen2018deep, oord2018representation}.

Unsupervised learning not only reduces the need for labeled datasets, it also makes it possible to build comprehensive hierarchical representations that provide a deep insight into the nature of the input data. This is particularly important in speech compression, where efficiency depends on the completeness and compactness of the representation, which should capture all sorts of speech attributes. Yet despite the great potential of unsupervised learning, domain-specific representation learning, which can only capture a subset of the attributes from labeled data, is still prevalent in the literature. Currently, one of the very few approaches to extract comprehensive speech representations is the Vector Quantized Variational Autoencoder (VQ-VAE) \cite{oord2017neural}. Its use in recent deep learning-based speech coders and synthesizers \cite{garbacea2019low, casebeer2021enhancing, wang2019vector} substantiates the need for compact and complete speech representations.

In this paper, we propose and evaluate a new approach for unsupervised learning and extraction of speech representations that heavily relies on the principles of cognition. First, a two-stage neural network model is used to extract representations in two levels of abstraction, with a lower stage and an upper stage processing information from short and long frames of data, respectively. Secondly, a top-down pathway between stages is introduced, which has the effect of improving the quality of the representations. Finally, predictive coding is used as the learning strategy. The performance of the proposed approach is measured in terms of classification accuracy for speaker identity, emotions and phonemes. To position the results of the proposed approach with respect to the current state of the art, Contrastive Predictive Coding (CPC) \cite{oord2018representation} is used as a baseline. We observe that the second stage of the proposed model delivers a compact and remarkably high-quality long-term representation of the speech signal. The quality of the short-term representation extracted by the first stage is improved compared to that of the CPC baseline, especially when the dimension of the representation is reduced. Finally, we demonstrate that the extracted representations are extremely robust to quantization.

\begin{figure*}
\centering
  \includegraphics[width=16.5cm]{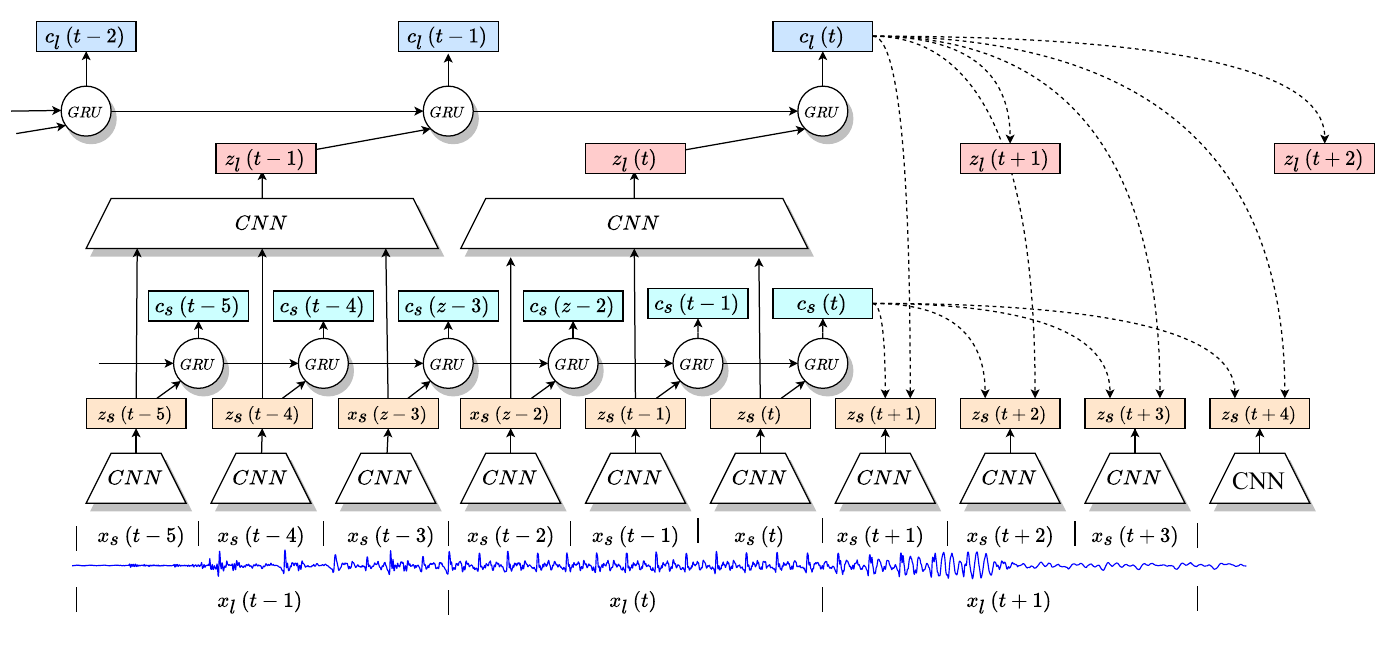}
  \caption{Representation of the architecture and learning algorithm of the cognitive coding model.  The ratio between long and short frames in the diagram is chosen to be three for purpose of the demonstration. In this study the actual frame ratio is eight.}
\label{fig:algo}
\end{figure*}

\section{Relation to prior work}
\label{sec:related}

The proposed Cognitive Coding model utilizes predictive coding in two stages and includes a top-down process between stages. These two stages produce two representations that evolve at a different pace and thus correspond to different levels of abstraction. The representations are extracted by maximizing the mutual information between the latent variables and the speech signal. Finally, the mutual information is maximized by minimizing a contrastive loss.




Mutual information is a fundamental quantity measuring the relationship between random variables. In previous work, it has been used in the formulations of Generative Adversarial Networks (GANs) \cite{goodfellow2014generative} and Variational Autoencoders (VAEs) to make them learn interpretable representation of data \cite{chen2016infogan,alemi2016deep,chorowski2019unsupervised}. Noise Contrastive Estimation (NCE) is a method for parameter estimation of probabilistic models by discriminating data from noise \cite{gutmann2010noise,hyvarinen2016unsupervised}. In the model called Contrastive Predictive Coding (CPC) \cite{oord2018representation}, NCE is also formulated as a probabilistic contrastive loss that maximizes the mutual information between the encoded representations and the input data.

In the CPC model, an encoder maps the input data to a sequence of latent variables, and an autoregressive model produces another sequence of latent variables. The InfoNCE loss introduced in \cite{oord2018representation} optimizes the discrimination of a positive sample from multiple negative samples. In this paper, we optimize a similar objective with consideration of two levels of abstraction and the presence of a top-down process. We also implemented the CPC algorithm as a baseline against which to compare our results.





\begin{figure*}
\begin{minipage}[b]{.24\linewidth}
  \centering
  \centerline{\includegraphics[width=3.9cm]{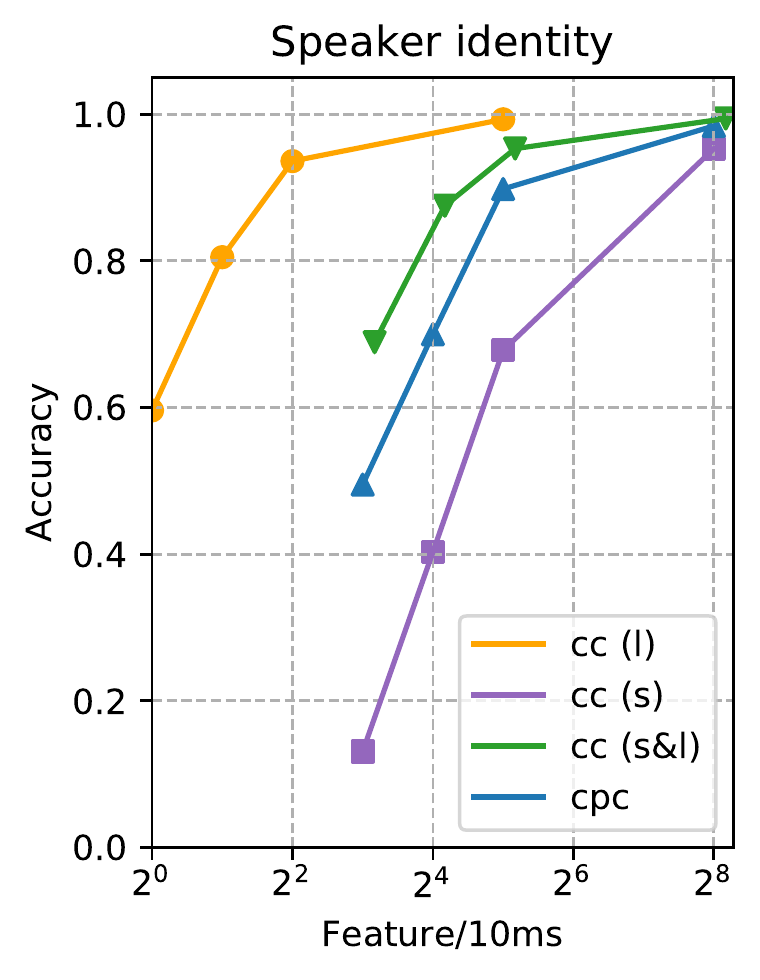}}
  \centerline{(a)}\medskip
\end{minipage}
\begin{minipage}[b]{.24\linewidth}
  \centering
  \centerline{\includegraphics[width=3.9cm]{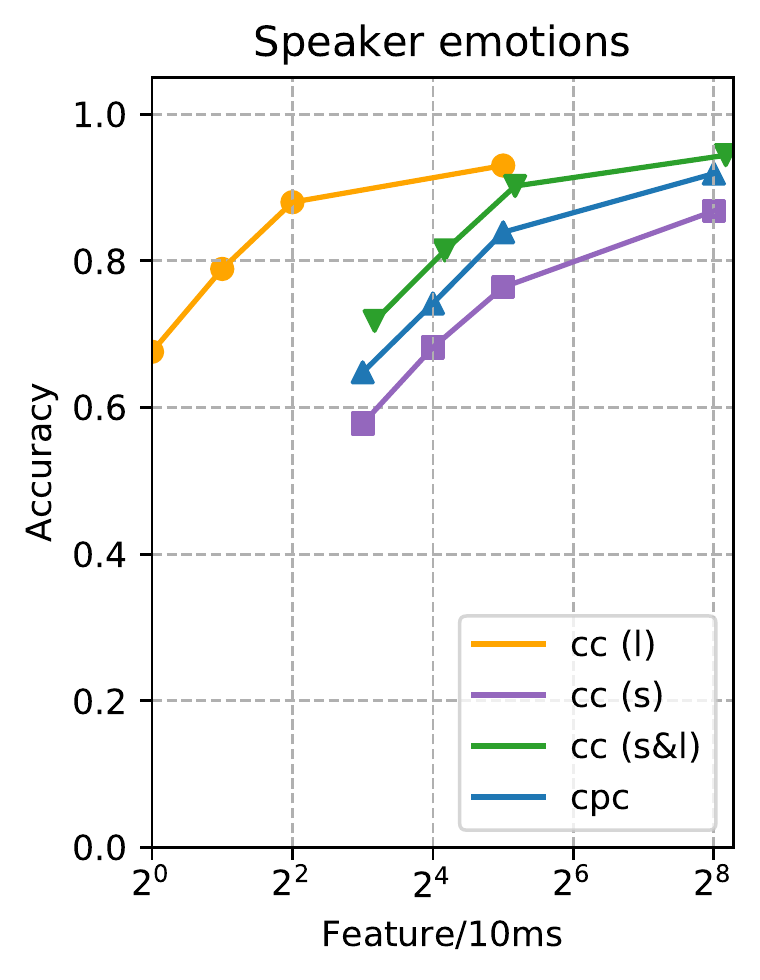}}
  \centerline{(b)}\medskip
\end{minipage}
\begin{minipage}[b]{0.24\linewidth}
  \centering
  \centerline{\includegraphics[width=3.9cm]{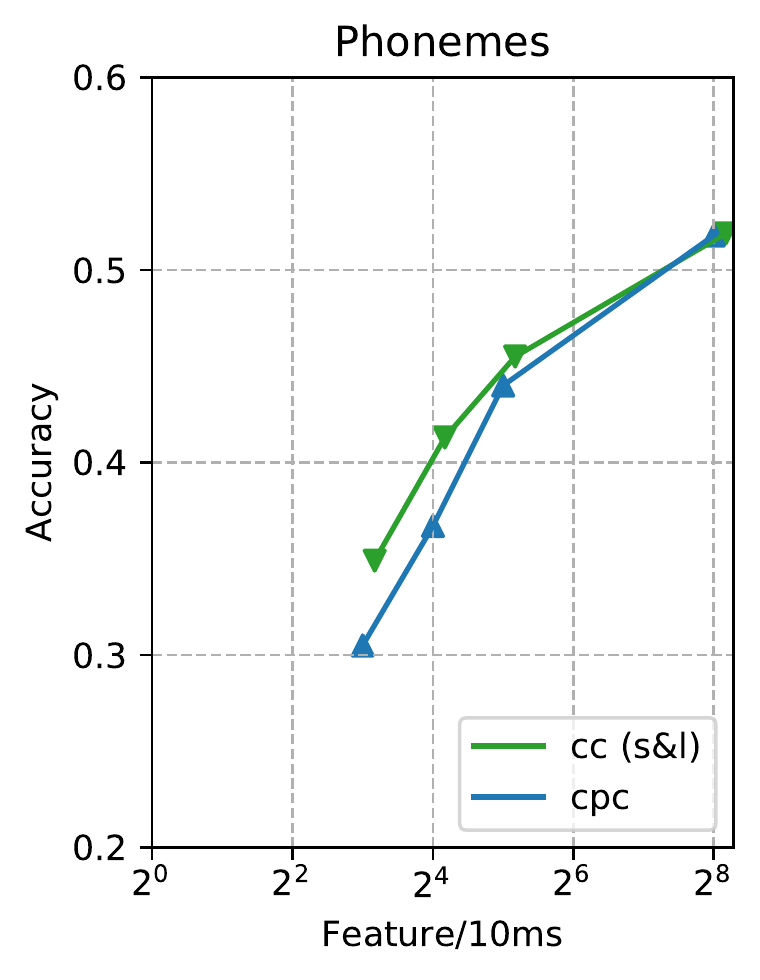}}
  \centerline{(c)}\medskip
\end{minipage}
\begin{minipage}[b]{0.24\linewidth}
  \centering
  \centerline{\includegraphics[width=3.9cm]{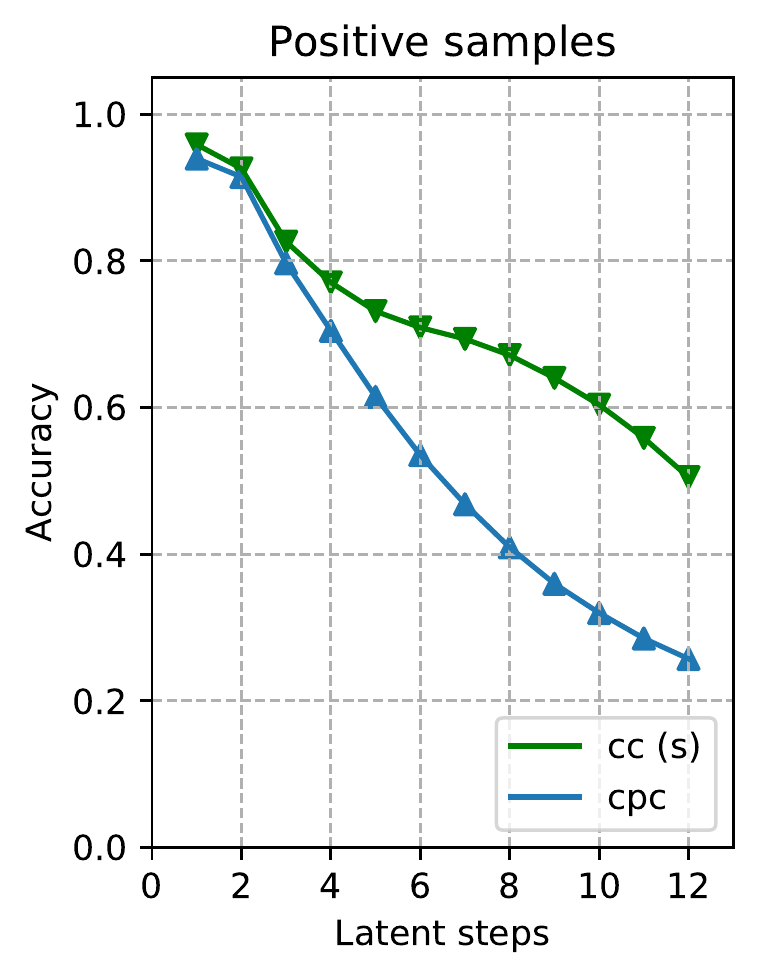}}
  \centerline{(d)}\medskip
\end{minipage}

  \caption{ Linear classification of attributes and prediction accuracy of positive samples in the loss function. (s: short-term, l: long-term, CC: Cognitive Coding. CPC: Contrastive Predictive Coding.) }
  \label{fig:accs}
\end{figure*}

\section{cognitive coding of speech}
\label{sec:model}

The architecture and learning algorithm of the Cognitive Coding model are illustrated in Fig.\ref{fig:algo}. The architecture can be described as follows. First, an encoder maps short frames of speech signal $x_{s}(t)$ to a sequence of latent variables $z_{s}(t)$ while decreasing the temporal resolution. Then, another encoder maps the first sequence of latent variables $z_{s}(t)$ to another set of latent variables  $z_{l}(t)$ while further decreasing the temporal resolution and increasing the receptive field to match long frames of speech signal. In this study, we use layers of Convolutional Neural Networks (CNNs) as encoders. Finally, two autoregressive models map $z_{s}(t)$ and $z_{l}(t)$ to two sequences of contextual representations $c_{s}(t)$ and $c_{l}(t)$.  In this study we use Gated Recurrent Units (GRUs) for the autoregressive models.


We begin by describing the learning algorithm for the lower stage of the model. In this lower stage, the mutual information between both contextual representations and short frames of speech signal can be expressed as:

\begin{equation}\label{eq:1}
I(x_{s};c_{s},c_{l})=\sum_{x_{s},c_{s},c_{l}}p(x_{s},c_{s},c_{l})log\frac{p(x_{s}|c_{s},c_{l})}{p(x_{s})}
\end{equation}

The following unnormalized density ratio captures the mutual information between a future short frame of speech signal at step  $t+k$ and both contextual representations:

\begin{equation}\label{eq:2}
f_{k}(x_{s}(t+k),c_{s}(t),c_{l}(t))\propto\frac{p(x_{s}(t+k)|c_{s}(t),c_{l}(t))}{p(x_{s}(t+k))}
\end{equation}

As in the CPC model, we do not use a generative model to produce future frames of speech signal. Rather, we use the following quantity to approximate $f_{k}$:
\begin{equation}\label{eq:3}
\exp (z_{s}^{T}(t+k)W_{s}(k)g(c_{s}(t),c_{l}(t)))
\end{equation}
In equation (\ref{eq:3}), $W_{s}(k)$ is a linear transformation used for the prediction of $z_{s}(t+k)$ ($k$ steps in the future) and  $g(c_{s}(t),c_{l}(t))$ is a function of both contextual representations that constitutes the input of the linear transformation. While a neural network could be used for $g$ to perform a nonlinear transformation, we simply repeat the long-term representation to match the temporal resolution of the short-term representation and concatenate it with the short-term representation to be used as input for the linear prediction of $z_{s}(t+k)$ by $W_{s}(k)$ . This is perfectly justified because the upper stage of our model produces a long-term representation that is easily interpretable by linear classifiers (see section \ref{sec:linear}).

Finally, the loss function is derived according to noise contrastive estimation which is the categorical cross entropy of classifying one positive sample of short frames of speech signal from $N-1$ negative ones:
\begin{equation}\label{eq:4}
L_{N}=\mathop{\mathbb{E}}_{X_{s}}\left [ log\frac{f_{k}(x_{s}(t+k),c_{s}(t),c_{l}(t))}{\sum _{x_{s}(j)\in X_{s}}f_{k}(x_{s}(j),c_{s}(t),c_{l}(t))} \right ]
\end{equation}

For the upper stage of the model, an equivalent of equations (\ref{eq:1}-\ref{eq:4}) can be derived based on long frames of speech signal $x_{l}(t)$.  $c_{s}$ is omitted from equations (\ref{eq:1}-\ref{eq:2}). Furthermore, since there is no top-down pathway in the upper stage, the prediction of $z_{l}(t+k)$ is based only on the long-term contextual representation $c_{l}(t)$ and the approximation for the density ratio becomes:
\begin{equation}\label{eq:5}
\exp (z_{l}^{T}(t+k)W_{l}(k),c_{l}(t))
\end{equation}

The loss function is derived by substituting equation (\ref{eq:5}) in equation (\ref{eq:4}), and samples are drawn from long frames of speech signal.

\section{EXPERIMENTS}
\label{sec:experiments}

This section presents experimental results regarding various speech attributes and investigates the effects of dimensionality reduction and quantization on the quality of the representations. Two different datasets were used. First, a 100-hour subset of the LibriSpeech dataset \cite{panayotov2015librispeech} was used to evaluate the performance of the proposed approach on phonemes (a short-term attribute) and on speaker identity (a long-term attribute). We used forced-aligned phoneme labels as well as the test and train split from \cite{oord2018representation} so that we could obtain comparable results. Secondly, we used the Emov-DB dataset \cite{adigwe2018emotional}  to evaluate the performance of the proposed approach on speaker emotions which is another long-term attribute.

The encoder used in the lower stage consists of five layers of CNNs with filter sizes [10, 8, 4, 4, 4] and with strides [5, 4, 2, 2, 2]. The encoder in the upper stage consists of three layers of CNNs with filter sizes [4, 4, 4] and with strides [2, 2, 2]. Each layer has 512 hidden units with ReLu activations. As a result, the lower and upper encoders downsample their input by a factor of 160 and 8, respectively. We trained on 20480-sample windows of speech signal sampled at 16kHz. As a result, the lower and upper encoders produce $z_{c}$ and $z_{l}$ vectors of features once every 10ms and 80ms, respectively. We decided that the dimension of the hidden state of GRUs would be either 8, 16, 32 or 256 so that the network can produce representations of various dimensions. Prediction is done twelve steps in the future, which extends the window of prediction up to 120ms in the future for the lower stage and 960ms for the upper stage. We trained with a learning rate of 2e-4, using mini batches of 8 samples, and performed approximately 300k updates.

\subsection{Linear classification}
\label{sec:linear}

The performance of our model is measured by training linear classifiers for various speech attributes to show to what extent the extracted features are linearly interpretable. Fig.\ref{fig:accs} (a-c) presents the performance of linear classification for speaker identity, emotion and phonemes.  Fig.\ref{fig:accs} (d) shows the ability of the lower stage of the proposed model to predict positive samples in the loss function up to twelve steps in the future. The results are reported for classifying contextual representations extracted from long frames of signal (l), short frames of signal (s), combined contextual representations (s{\&}l) as well as contextual representations of the CPC model. The following observations can be made based on the results.

Regarding the baseline, the results reported in \cite{oord2018representation} for the 256-dimension representation which produces 256 features every 10ms are 97.4\% and 64.6\% of accuracy for speaker identity classification and phoneme classification, respectively. With our implementation of CPC, we were able to achieve a higher accuracy of 98.4\% for speaker identity but a lower accuracy of 51.9\% for phonemes.

Since the upper stage of our model produces a set of features for each 80ms of speech signal, the number of features per 10ms is 8 times less relative to the lower stage of our model and to the CPC model. For long-term attributes (speaker identity and emotion) the proposed network outperforms CPC in terms of linear classification for combined 256-dimension representations by achieving an accuracy of 99.3\% and 94.4\% for   speaker identity and emotion, respectively. The corresponding accuracy achieved by the CPC model was 98.4\% and 91.9\%. By reducing the dimensionality of the representations, we observe that a high degree of linear separation between speaker identities and emotions is maintained when considering the features extracted by the upper stage of our model. Features extracted by the lower stage provide lower performance for long-term attributes. Overall this is a desirable effect that we attribute to the top-down pathway which provides a link to predict long-term attributes that are present in a short frame of signal.

Regarding linear classification of phonemes based on contextual representations, we achieved 52\% accuray, a lower performance compared to the state of the art with the forced aligned features provided by \cite{oord2018representation} and this is true even with our implementation of CPC baseline model. However, phoneme information is encoded in latent variable $z_{s}$ which has a smaller receptive field compared to both contextual representations. Besides, not all information is linearly interpretable. In an experiment we used a classifier with one hidden layer on contextual representations and latent variables $z_{s}$ and $z_{l}$ and accuracy increased to 64.1\%. Features of  $z_{s}$ are also a candidate for dimentionality reduction to encode information in smaller time scale.

We also investigated the effect of the top-down pathway on the prediction of positive samples in the lower stage and compared the performance of our model with that of the CPC baseline in the same setup. Fig \ref{fig:accs} (d) shows that the proposed approach is able to predict positive samples of short frames more efficiently beyond 3 latent steps.

\subsection{Quantization}

\begin{figure}

\begin{minipage}[b]{.32\linewidth}
  \centering
  \centerline{\includegraphics[width=2.7cm]{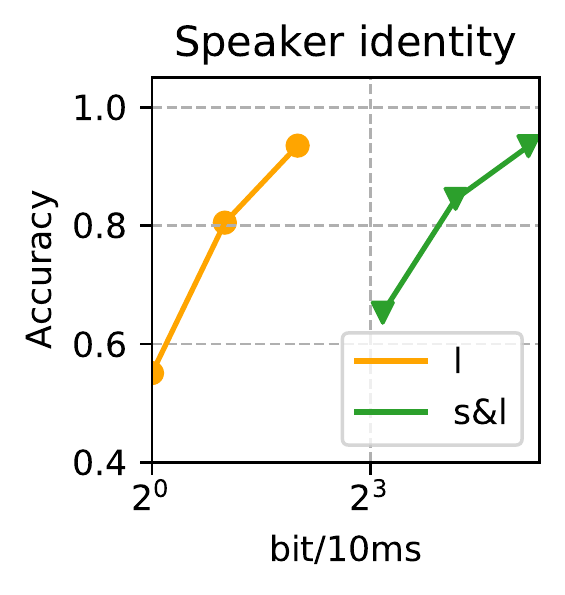}}
\end{minipage}
\begin{minipage}[b]{.32\linewidth}
  \centering
  \centerline{\includegraphics[width=2.7cm]{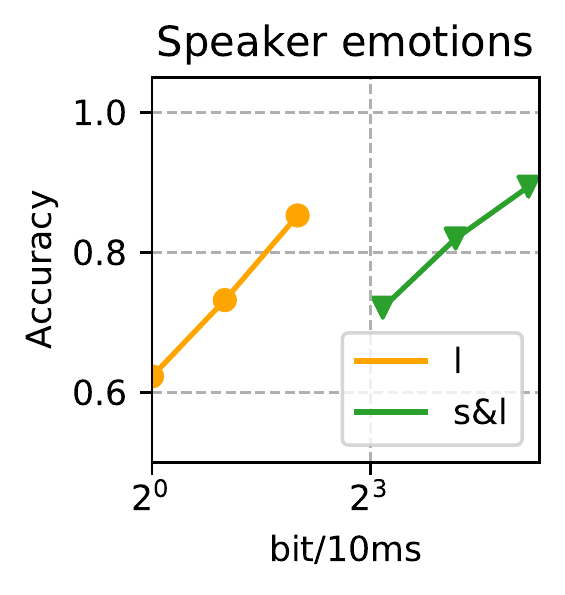}}
\end{minipage}
\hfill
\begin{minipage}[b]{.32\linewidth}
  \centering
  \centerline{\includegraphics[width=2.7cm]{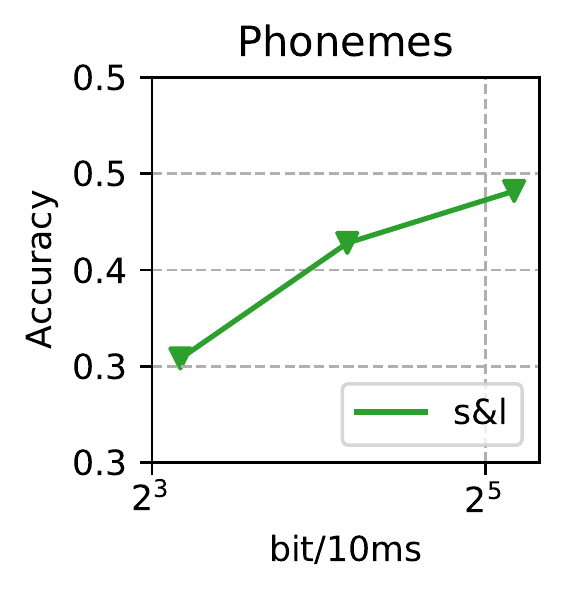}}
\end{minipage}
\caption{Linear classification of quantized features.}
\label{fig:quant}
\end{figure}

In this study, we also investigated the compressibility of the features. Since each stage predicts twelve time steps in the future, the contextual representations have a slow-evolving nature and we observe that the features exhibit a high degree of temporal dependency. For this reason, we decided we would quantize the features using 1-bit $\Delta$-modulation. The initial values of the features are encoded on 5 bits. Fig.\ref{fig:quant} shows the results obtained when the features are quantized for the most interesting configurations from Fig.\ref{fig:accs}. We only consider representations with 32 dimensions and less because they are the most likely to be used in speech compression applications. For the majority of the cases, the performance of the linear classification is within 5\% of the corresponding performance from Fig.\ref{fig:accs}. Most notably, we observe that our model can encode long-term speech attributes such as speaker identity and emotion with more that 50\% accuracy at bitrates as low as 100 bit/s.

\section{conclusion}
\label{sec:conclusion}

In this paper, we presented a new model for cognitive coding of speech that combines several principles of cognition. Specifically: (1) it produces a hierarchy of representations that correspond to different levels of abstraction; (2) it uses the predictive coding principle; and (3) it includes a top-down pathway between levels of abstractions. The hierarchy of representations captures a wide variety of speech attributes over a broad range of time scales. Experiments show that this hierarchy is also easily interpretable, well suited for compression, and remarkably robust to quantization. This cognitive coding model should therefore find applications in high-quality speech synthesis, voice transformation and speech compression.\looseness=-1

\bibliographystyle{IEEEbib}
\bibliography{refs}

\end{document}